\documentclass{aastex}
\usepackage{emulateapj5,graphicx}

\shorttitle{Metal-Poor Halo Stars and Reionization}
\shortauthors{Venkatesan}

\begin{document}

\title{A Cosmic Milestone: Constraints from Metal-Poor Halo Stars \\ on
  the Cosmological Reionization Epoch}
\author{Aparna Venkatesan\altaffilmark{1}} 
\affil{CASA, Department of Astrophysical and Planetary Sciences, \\ UCB 389,
University of Colorado, Boulder, CO 80309-0389}
\altaffiltext{1}{NSF Astronomy and Astrophysics Postdoctoral Fellow}
\email{aparna@casa.colorado.edu} 

\begin{abstract}

Theoretical studies and current observations of the high-redshift
intergalactic medium (IGM) indicate that at least two cosmic transitions
occur by the time the universe reaches gas metallicities of about $10^{-3}
Z_\odot$.  These are the cosmological reionization of the IGM, and the
transition from a primordial to present-day mode of star formation. We
quantify this relation through new calculations of the ionizing radiation
produced in association with the elements carbon, oxygen and silicon
observed in Galactic metal-poor halo stars, which are likely
second-generation objects formed in the wake of primordial supernovae. We
demonstrate that sufficient ionizing photons per baryon are created by
enrichment levels of [Fe/H] $\sim$ -3 in the environment of metal-poor halo
stars to provide the optical depth in the cosmic microwave background of
$\sim$ 0.1 detected by $WMAP$. We show, on a star by star basis, that a
genuine cosmic milestone in IGM ionization and star formation mode occurred
at metallicities of $10^{-4}$--$10^{-3} Z_\odot$ in these halo stars. This
provides an important link in the chain of evidence for metal-free first
stars having dominated the process of reionization by $z \sim$ 6.  We
conclude that many of the Fe-poor halo stars formed close to the end of or
soon after cosmological reionization, making them the ideal probe of the
physical conditions under which the transition from first- to
second-generation star formation happened in primordial galaxies.

\end{abstract}


\section{Introduction}

Theoretical studies of the early universe predict the existence of
first-generation stars which lack metals and consequently have unique
physical structures and properties \citep{bkl, tsv, sch02}.  Such objects
have not, however, been detected to date in either the high-redshift or
local universe. Therefore, indirect inferences from modelling the impact of
primordial stars and supernovae (SNe) on their environment through
radiative and chemical feedback remain important. The reionization of the
intergalactic medium (IGM) and the widespread metal enrichment of high-$z$
galaxies, QSOs and the IGM provide strong observational constraints for
such theoretical studies. A particularly effective tool in constraining the
masses and properties of the first stars has been to combine their ionizing
photon and nucleosynthetic signatures \citep{venktruran, tvs04}.

When we examine the current data and the results of theoretical work, an
interesting coincidence becomes apparent: that three important cosmic
transitions occur at gas metallicities of about $10^{-5}$--$10^{-3}
Z_\odot$. To begin with, the universe first generates sufficient ionizing
photons for hydrogen reionization when the IGM reaches these metallicities
\citep{mirrees97,venktruran}.  Second, theoretical studies of the cooling
and fragmentation of gas in early galaxies reveal that the primordial
stellar initial mass function (IMF) evolves from its metal-free to
present-day form at gas transition metallicities of $Z_{\rm tr} \sim
10^{-5}$--$10^{-3} Z_\odot$ \citep{brommetal01,schneider02,omukai05,
santoroshull}. This transition results from the physics of gas cooling in
early galaxies, when the dominant cooling channel switches from molecular
hydrogen to metal lines or dust grains, with the latter pathway dominating
in systems with $Z \la 10^{-4} Z_\odot$ \citep{omukai05}.  Third, there are
indications that the relative abundances of CNO and alpha-elements in
metal-poor stars in the Galactic halo have increased scatter and altered
ratios when the iron content of these stars falls below $\sim 10^{-3}$ of
solar values (\citealt{barklem05,cayrel}; \citealt{beersaraa05}, and
references therein). These subsolar mass objects are thought to be
second-generation stars that formed from metal-poor fragmenting gas in the
shells of the very first SNe in early galaxies \citep{salvaterra04}.  Such
stars contain the fossil imprint of nucleosynthesis from the preceding
generations of metal-free stars, reflecting the ``prompt inventory'' of
metals \citep{qianwass02,wadavenk}.

When these three points are taken together, we see that several
cosmological milestones occured by metallicities of $\sim 10^{-3} Z_\odot$,
a relation that must result from the strong feedback between the buildup of
ionizing radiation and the formation of second-generation stars.  In this
Letter, we quantitatively demonstrate a novel connection in the environment
of extremely metal-poor (EMP) halo stars between two of these
relationships: that the stellar population that was important for early
nucleosynthesis is also responsible for cosmological reionization. We
define EMP stars as those with [Fe/H] $\la$ -3 \citep{beersaraa05}. We
examine the significance of the second transition in relation to the other
two in a subsequent work, where we plan to use numerical simulations to
investigate the spatially varying feedback between gas cooling and the
propagation of ionizing radiation in the first galaxies.

\section{Stellar Abundances and Ionizing Efficiency}

We begin by showing the trends of element abundances in EMP stars from
current observations. We follow our approach in \citet{vns06} and take the
data from \citet{cayrel}, and \citet{tvs04} and references therein. For the
two most iron-poor stars, HE-0107-5240 and HE-1327-2326, we use the
published abundances from \citet{christlieb04}, \citet{aoki04},
\citet{frebel05} and \citet{frebel06}. The last of these papers has
recently revised the CNO abundances of HE-1327-2326 when 3D NLTE
corrections are taken into consideration. We focus on the elements C, O, Si
and Fe, as these are amongst the dominant metals created in primordial SNe
and are the most important coolants in metal-free gas
\citep{brommloeb03,santoroshull}.

The trends of [C/H], [O/H], [Si/H] and [C/O] as a function of iron
abundance [Fe/H] are shown in Figure 1, where for an element X, we define
[X/H] as the ratio of the measured column densities of X/H to its solar
ratio. We display those points with [Fe/H] $\la$ -2 in order to span the
predicted $Z_{\rm tr}$ in individual elements
\citep{brommloeb03,santoroshull}, as we will see below.  We assume that EMP
stars with [Fe/H] $\la$ -3 have been enriched only by metal-free stars,
consistent with the estimate of $Z_{\rm tr}$ from \citet{santoroshull} and
with recent calculations using merger tree models in which stars of
metallicities $\la$ $10^{-3} Z_\odot$ reflect enrichment from only one to a
few primordial SNe \citep{tum06}.  \\

\centerline{\epsfxsize=1.0\hsize{\epsfbox{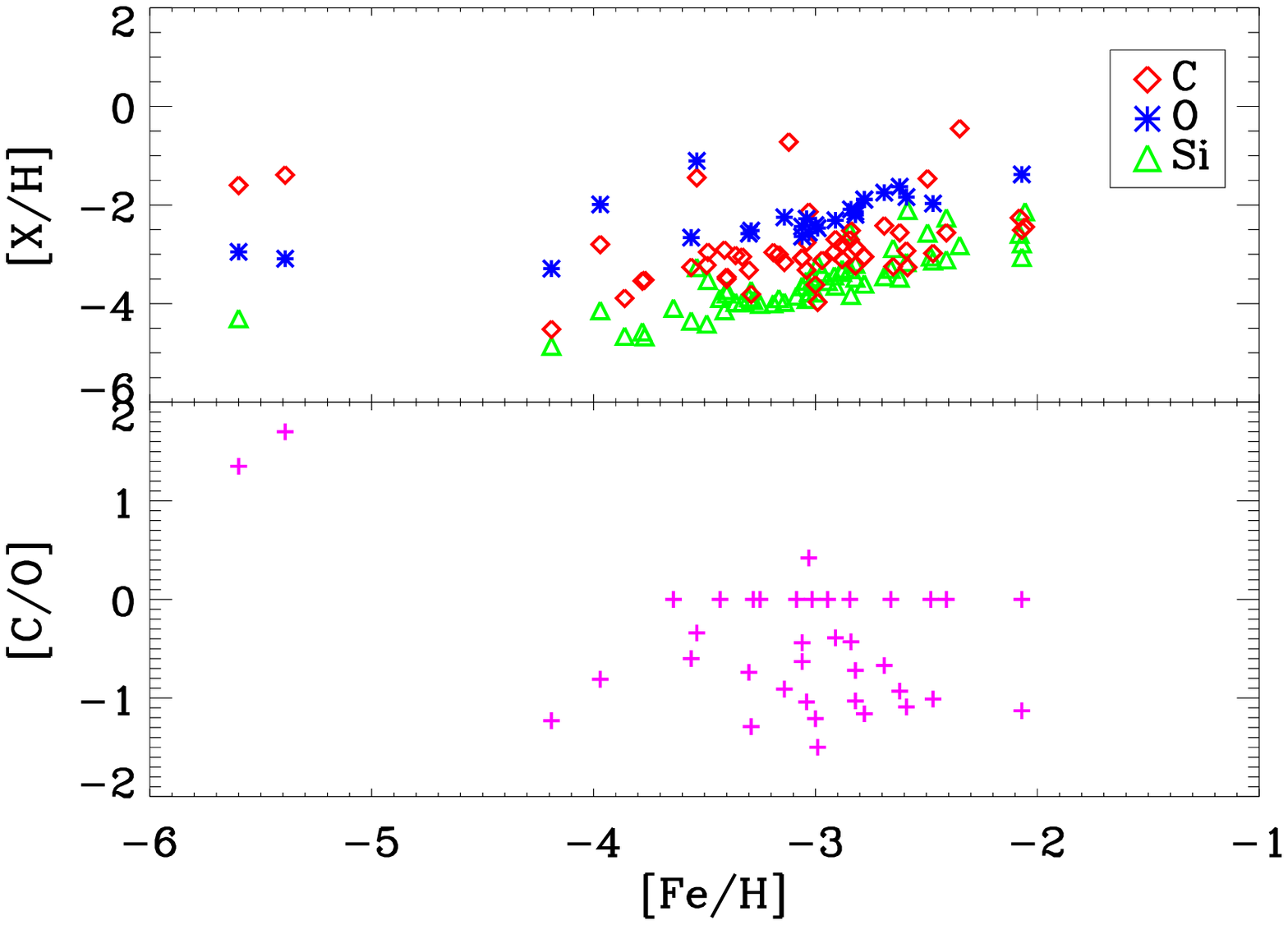}}}

\vspace{0.1in} 
\figcaption{The current data from Galactic metal-poor halo stars on the
  abundances of C, O and Si relative to H, and of [C/O], as a function of [Fe/H].
}
\vspace{0.2in}

Figure 1 reveals some interesting trends. First, for those stars that have
both abundances measured, [C/O] shows no clear trend with [Fe/H]; in fact,
it appears to have its maximum value at the lowest [Fe/H]. Such element
abundance ratios are often used as a cosmic clock to track the creation of
elements from stars of differing lifetimes, but such a relation may not be
easily derived in EMP stars, a point which we return to in the discussion.
Second, carbon and silicon roughly track Fe in enrichment relative to H,
whereas oxygen does not. Again, the exceptions to this trend are the most
Fe-poor stars. Note that some of the stellar data shown here come from
surveys that select against C-rich EMP stars \citep{cayrel}, owing to the
complicated analyses involved in such cases. The more recent survey by
\citet{barklem05} does not have this bias, although these authors focus
mostly on $r$-process element abundances in EMP stars.

A direct translation of the observed abundances in EMP halo stars to yields
from primordial SNe and constraints on the first-stars IMF would be ideal,
but is complicated by several factors. Mass transfer from a companion may
alter original abundance ratios, but is difficult to model owing to the
unknown binary fraction of the first stellar generations, and the role of
AGB phase dredge-ups and mass loss for metal-free stars. We have not
plotted those EMP stars known to be in binaries in Figure 1, but it is
possible that many of these stars have as yet undetected companions. A
uniform sample coverage of the CNO elements would constrain this scenario
better.

In order to assess the implications of the nucleosynthetic data in EMP
stars for reionization, we follow the approach in \citet{madshull} and
\citet{venktruran}. We calculate the ionizing efficiency, $\eta_{\rm Lyc}$,
which is essentially a conversion factor defined as the ratio of
the energy produced in rest mass of metals,
$M_Z c^2$, and the energy released in the H-ionizing continuum:
\begin{equation}
\eta_{\rm Lyc} \equiv E (h\nu \geq 13.6 \, {\rm eV})/(M_Z c^2) 
\end{equation}

The first stars are expected to be metal-free in composition, a factor that
critically influences their structure and leads to the emission of
significantly harder ionizing radiation relative to low-$Z$ stars
\citep{bkl, tsv, sch02}.  \citet{venktruran} showed that $\eta_{\rm Lyc}$
is a strong function of stellar metallicity and increases by a factor of
10--20 from $Z=Z_\odot$ to $Z=0$ stars, owing to the boosted ionizing flux
and substantially reduced metal yield from massive stars with decreasing
stellar metallicity. The latter trend is a result of the more compact
structure of metal-poor stars, which directly leads to greater element
fallback onto the remnant after the SN explosion.

The stellar IMF is assumed to have the form, $\phi(M) = \phi_0
M^{-\alpha}$, which is normalized over the mass range as, $\int dM M
\phi(M) = 1 $. We consider two primordial IMFs, one that has 100--1000
$M_\odot$ stars with a flat IMF slope (hereafter very massive stars; VMSs),
as some theoretical studies indicate \citep{omukai03,bromm02,naka}, and a
present-day IMF with 1--100 $M_\odot$ stars where the IMF slope has a
Salpeter value, $\alpha=2.35$.  We use the ionizing spectra from
\citet{bkl} and \citet{sch02} for VMSs, and the results from \citet{tsv}
for the 1--100 $M_\odot$ IMF.  We take the metal yields from \citet{ww95}
for 10--40 M$_\odot$ stars, which end their lives as Type II SNe, and from
\citet{hw02} for stars with masses 140--260 $M_\odot$ which explode
entirely as pair-instability SNe. All other massive stars are assumed to
collapse into black holes with no contribution to the nucleosynthetic
output. Varying the upper and lower mass limits in the VMS IMF will not
significantly impact our conclusions, owing to the narrow mass range of
element synthesis from VMS SNe, and to the near-identical ionizing photon
production per stellar baryon in all stars with masses $\ga 300 M_\odot$
\citep{bkl}.

We do not include cases where the conversion efficiency is derived
including the metal yield from intermediate-mass stars, as $\eta_{\rm Lyc}$
remains approximately the same for oxygen and exactly the same for Si. For
carbon, $\eta_{\rm Lyc}$ decreases by a factor of a few with
intermediate-mass stars owing to their large yield in C. For the purposes
of this paper, we wish to track the growth of ionizing radiation which is
dominated by the massive stars in the IMF. We do not include the yield from
Type Ia SNe in our calculations.  Recent studies indicate that such SNe may
not occur in metal-free stellar populations \citep{kobayashi}, although
this is a sensitive function of the SN mechanism itself and of the unknown
binarity in $Z=0$ stars.

As shown in \citet{venktruran} and \citet{mirrees97}, calculations of
$\eta_{\rm Lyc}$ can be used to compute the number of ionizing photons
per baryon generated in association with a metallicity that is observed in
a given system. These papers focussed on the IGM, and pointed out that the
universe first generates about 10 ionizing photons per IGM baryon for the
observed IGM metallicity of $\sim 10^{-2.5} Z_\odot$. This estimate is
however subject to the uncertainties of star formation efficiency and the
escape fraction of ionizing radiation from early galaxies. Here, we examine
the ionizing efficiency in the environment of EMPs, and calculate
$N_{\gamma}/N_{\rm b}$, the number of ionizing photos created per baryon in
stars, which is not related to the above highly uncertain astrophysical
parameters. Following \citet{venktruran}, for an element $i$,
$N_{\gamma,i}/N_{\rm b} = \eta_{\rm Lyc,i} \times Z_i \times$ (1 GeV/$\langle
E_{\rm Lyc} \rangle)$; we assume the IMF-averaged energy of a Ly-continuum
photon, $\langle E_{\rm Lyc} \rangle$ = 27 eV (30 eV) for metal-free stars
in a present-day (VMS) IMF.

\section{Results and Discussion}

For our calculations here, we take the values of $\eta_{\rm Lyc}$ for
carbon and oxygen computed in \citet{venktruran}, and compute them
independently for silicon. We find that for the 1--100 $M_\odot$ 
Salpeter-slope IMF, $\eta_{\rm Lyc}$ has values of 0.48, 0.2 and 1.34 for
C, O and Si respectively; for a VMS IMF, the corresponding values are
0.098, 0.01, and 0.057. Using these numbers, we show the results of our
calculations in Figure 2, where the number of ionizing photons created per
baryon in EMP stars, $N_{\gamma}/N_{\rm b}$, is shown as a function of
[Fe/H] for the elements C, O and Si in three separate panels. In each case,
the ionizing photon contribution from a present-day and a VMS IMF are
plotted.  

A number of trends seen in Figure 2 are of relevance to the first-stars
field. First, VMS make a low but persistent contribution to the overall
ionizing photon budget in association with the metals locked in EMP stars,
typically an order of magnitude less than the 1--100 $M_\odot$ IMF. This is
because VMSs are in general less efficient at generating total ionizing
radiation per unit metal yield relative to other metal-free stellar
populations, owing to the large metal production from VMSs.  Second, the
slow buildup of metals and the associated ionizing radiation is apparent,
for each element and for each IMF. This trend appears most obvious for
oxygen, although there is a significant scatter in the correlation between
$N_{\gamma}/N_{\rm b}$ and [Fe/H] at any given [Fe/H]. As we stated
earlier, the notable exceptions are the two stars at the lowest values of
[Fe/H]. 

Third, we show in each panel the approximate criterion to generate an
optical depth in the cosmic microwave background (CMB) of about 0.1,
consistent with data from the {\it Wilkinson Microwave Anisotropy Probe}
($WMAP$; \citealt{spergel06}), of about $N_\gamma/N_{\rm b} \sim$ 34,000
\citep{tum06,tvs04}.  This number is derived from interpolating between
cases studied by \citet{tvs04} where the lifetime-integrated ionizing
photon production from various stellar populations and IMFs were calculated
and used as inputs in detailed cosmological reionization
models. Interestingly, the EMP data indicates that this line is crossed at
least once {\it and perhaps twice} between -4 $\leq$ [Fe/H] $\leq$ -3 by a
present-day IMF. This is best seen in the panel corresponding to Si, where
a clear dip in $N_\gamma/N_{\rm b}$ occurs between -3.6 $\leq$ [Fe/H]
$\leq$ -3. The VMS IMF appears to never meet this criterion over the
metallicity ranges considered here, consistent with the role of VMSs in IGM
reionization predicted by \citet{venktruran}.  \\

\centerline{\epsfxsize=1.0\hsize{\epsfbox{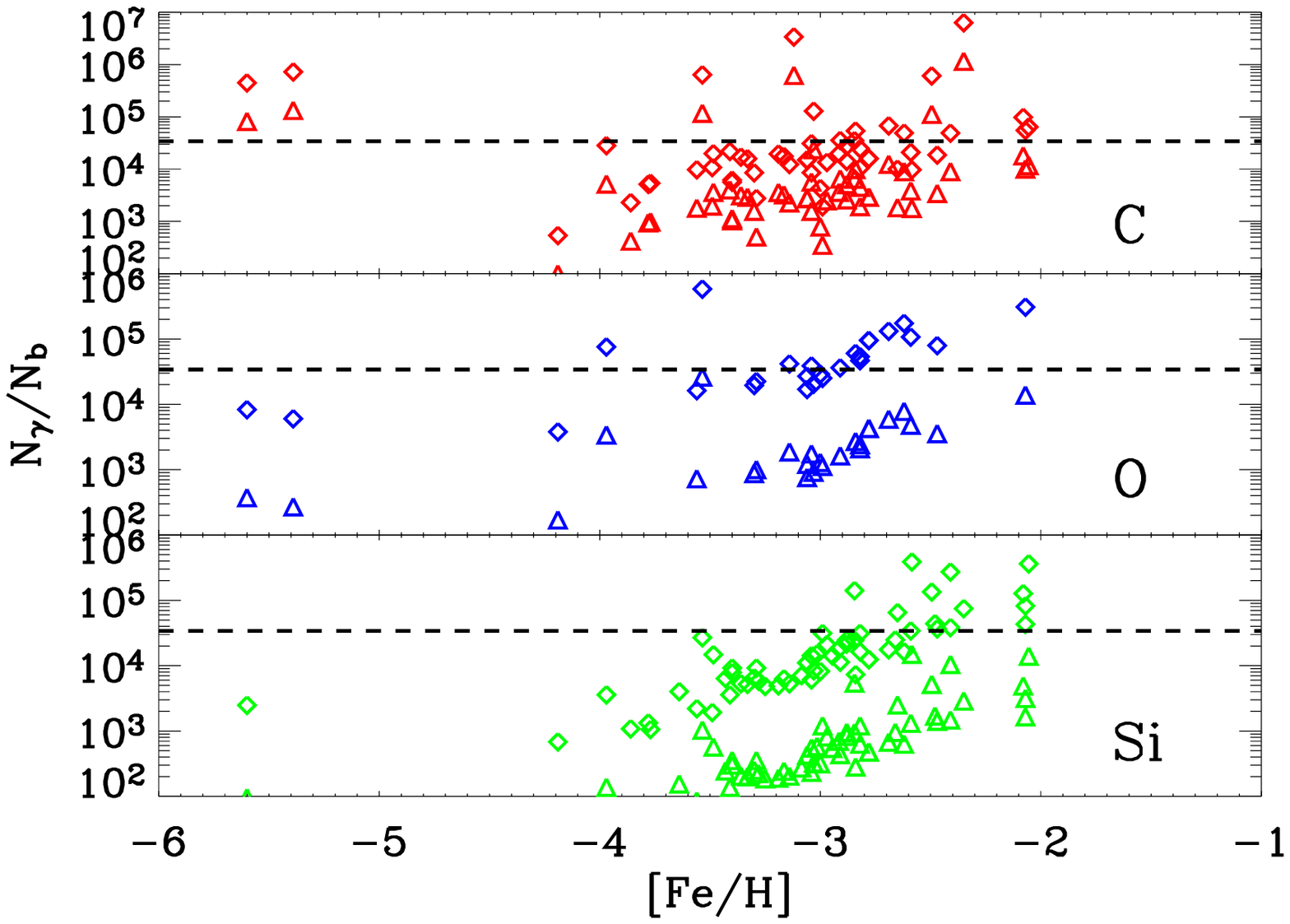}}}

\vspace{0.1in} 
\figcaption{The minimum number of ionizing photons per
baryon in stars that is created in association with specific metals
observed in MP halo stars. Top, middle and bottom panels correspond to the
elements C, O, and Si. The thick dashed black line in each panel
corresponds to the value required to roughly match the $WMAP$ data for an
optical depth of 10\% in the CMB. All cases are for metal-free
stars; each panel displays values for the 1--100 $M_\odot$ (diamond points)
and VMS (triangle points) IMFs considered in this work. See text for
discussion.}
\vspace{0.2in}

Conducting a metal census of the universe from EMP stars and from the IGM
is important in both environments, despite their differing systematics. The
IGM contains nearly all of the baryons at high redshifts but the former
case permits a detailed star-by-star analysis of the cosmic enrichment
history.  By working with nucleosynthetic data in EMP stars rather than the
IGM, we have not assigned a global fixed metallicity with the appropriately
scaled individual element values according to their solar ratios. Rather,
we have derived parallel constraints for reionization from a number of
elements independently, which is a considerably stronger approach. In
addition, observations of EMP stars relative to IGM absorbers provide
significantly larger element coverage in the well-constrained environment of
stellar atmospheres, a situation that will greatly improve in the near
future.

Although we naively apply the same factor for ionization efficiency for
[Fe/H] $\ga$ -3, $\eta_{\rm Lyc}$ is likely to be closer to the values
associated with metal-poor stars, typically a factor of a few to 10 lower,
owing to their increased metal yield. Depending on the individual metal
abundance, this could lead to a less steep rise or flattening in the
behavior of $N_{\gamma}/N_{\rm b}$ at [Fe/H] $\ga$ -3. This prediction is
somewhat more speculative, as the integrated contribution from many
subsequent stellar generations of varying metallicities will be difficult
to distinguish in this environment.

The notable exceptions to most of the trends noted above are the two most
Fe-poor stars at [Fe/H] $\sim$ -5.5. At such low metallicities, one would
expect that these stars reflect enrichment from a single SN
\citep{tum06}. These stars are, however, highly over-enriched in several
elements including C, O and Si, to the degree that they are actually not
metal-poor, but rather at metallicities of nearly a tenth solar. How these
stars came to acquire such skewed element abundances relative to solar
ratios remains puzzling at present.  Some possibilities include pollution
from a binary companion, from a subluminous hypernova with fine-tuned
ratios of mixing and fallback \citep{iwamoto05}, or the effects of stellar
rotation and ensuing mass loss which could dramatically boost the
production of CNO elements \citep{chiappini06}. The ratio of these elements
to iron or relative to each other is sometimes used as a chronometer in
other astrophysical environments. For the most Fe-poor stars, however, it
is clear that ratios such as [Fe/O] or [C/O] are not easily related to
time. These should normally increase with age of a stellar population as
the stars of progressively lower mass evolve off the main sequence, but as
we saw in Figure 1 for EMP stars, [C/O] achieves its highest values at the
lowest [Fe/H]. Perhaps these stars were created in atypical environments as
the IGM approached reionization, as seen from the generation of ionizing
radiation in Figure 2 -- the most Fe-poor stars appear to have formed at
epochs when, for a metal-free present-day IMF, $N_{\gamma}/N_{\rm b}$
nearly equals or exceeds the criterion from $WMAP$, at least for C. One
interesting explanation of this, in addition to those above, could be that
the most Fe-poor halo stars are also true second-generation stars: ones
that formed through the dust grain cooling pathway ($Z_{\rm tr} \sim
10^{-5} Z_\odot$) rather than through metal line cooling using C, O or Si
($Z_{\rm tr} \sim 10^{-3} Z_\odot$). The unusual element trends in these
stars could then be explained through the element segregation predicted in
a scenario involving dust transport in early galaxies. Here, conditions in
primordial SN remnants result in the radiative transport of dust in the hot
radiation field from metal-free stars, leading to selective element
enhancement from compounds such as graphites and silicates in the cooling
gas in SN shells \citep{vns06}, which are likely the sites of EMP star
formation.

What do the results in this paper imply for cosmology? Those EMP stars which
have observational age estimates are about 12--13 Gyr old
\citep{snedenetal03,beersaraa05}. In the context of
our results, it appears that most of the EMP stars formed close to the end
of or soon after cosmological reionization.  That reionization occurs in
the first Gyr of the universe (by $z \sim$ 6) is no surprise, as this is
already clear from Gunn-Peterson studies of the high-$z$ IGM \citep{becker}
and from current CMB data. A more interesting result of translating the EMP
star abundances in C, O and Si into the accompanying production of ionizing
photons per baryon in stars is the clear buildup of ionizing radiation
until reionization is accomplished at [Fe/H] $\sim$ -3, [Si/H] $\sim$ -3.5
and [O/H] $\sim$ -2.5 ([C/H] ranges from -4 to -2 at these
Fe-metallicities). At this point, the initial stellar population
(presumably metal-free) must turn off, as seen from the altered element
abundance patterns at higher [Fe/H], leading to a new pattern of
nucleosynthesis from different SNe. \citet{cayrel} argue for the presence
of a primordial SN population up to [Fe/H] or [Mg/H] values of about -3,
with a combination of primordial and nonprimordial SNe at higher
metallicities. Such a truncation in the mode of star formation is most
likely attributed to the critical metallicity being reached in the gas
in which EMP stars formed, signalling an end to the first-stars epoch.
Given certain conditions of density and temperature, $Z_{\rm tr}$ can be
higher in iron than in carbon, oxygen, or silicon. This is not consistent,
however, with the overall trend of enhanced C, O and Si at {\it low}
[Fe/H].  A naive projection of the above metallicities in Fe, Si, O and C
at which reionization ``occurs'' in Figure 2 implies that EMP stars should
have formed in gas of number densities $\ga 10^3$ cm$^{-3}$
\citep{santoroshull}.

A more puzzling coincidence is why the end of the first-stars phase should
occur at about the epochs corresponding to cosmological reionization. This
points towards the possibility of strong feedback between intrahalo metal
reincorporation timescales and halo-IGM transport of ionizing radiation. We
return to this issue in a subsequent work, where we will quantitatively
relate the reionization epoch to the transition metallicity in early
galaxies. Clearly, EMP stars provide an important window to study the
transition from first- to second-generation star formation and the physical
conditions at which this occurs in primordial galaxies. The novel result
derived here is that the abundances in such stars show a strong relation to
cosmological reionization. Targetted observations are required with more
uniform element coverage of EMP stars, particularly the C-rich stars, in
the critical range of -6 $\la$ [Fe/H] $\la$ -4. This range currently has
very little data, and will be useful in closing the loop in our
understanding of these important cosmic events at high redshifts.

\acknowledgements

I thank Tim Beers, Jason Tumlinson and an anonymous referee for helpful
comments on the manuscript.  I gratefully acknowledge the support of NSF
grant AST-0201670 through the NSF Astronomy and Astrophysics Postdoctoral
Fellowship program.

\end{document}